\documentclass[cits]{PoS}

\usepackage{graphicx}
\usepackage[figuresright]{rotating}

\newcommand{\ket}{\,\rangle}
\newcommand{\bra}{\langle \,}

\def\m{\mu}

\title{Hadronization in three meson channels at tau decays and e+e- cross-section}

\ShortTitle{Hadronization in three meson channels at tau decays and e+e- cross-section}

\author{\speaker{Pablo ROIG}%
         \thanks{I would like to thank all the members of the organizing committee for making such a good and enjoyable conference. I acknowledge discussions with O.Shekhovtsova and J.Trnka. This work has been supported in part by the EU MRTN-CT-2006-035482 (FLAVIAnet), by MICINN (Spain) under grant FPA2007-60323 and by the Spanish Consolider-Ingenio 2010 Program CPAN (CSD2007-00042).}\\
        IFIC (CSIC- Universidad de Valencia), INFN LNF Frascati \& Physik Department TUM Munich\\
        E-mail: \email{paroig@ific.uv.es,pablo.roig@ph.tum.de}}

\abstract{The hadronic form factors in semileptonic decays of the $\tau$ and low-energy $e^+e^-$ scattering into hadrons can be described accounting for the following fundamental features: the right chiral structure at low energies, the large-$N_C$ limit of QCD and the known asymptotic behaviour of the theory. All these characteristics are implemented within our approach to the accuracy and within the limits discussed below. Our results -improved with respect to those shown in the talk \footnote{They correspond to the ones presented in the V Working Group on Generators for Low-Energy Physics held in Frascati the 6th and 7th of April, 2009.}- are compared to data for $\tau\to3\pi\nu_\tau$. Our setup in the vector current has been checked (using $CVC$) with $e^+e^-\to KK\pi$ data. We predict the hadronic spectra in all $\tau\to KK\pi\nu_\tau$ charge modes.\\
Our hadronic matrix elements are being included in the TAUOLA library and may be useful for the experimental collaborations that are analysing $\tau$ decays (BaBar and Belle) and to those that will join this effort in the future (BES and super-B factories). It might also serve for the ATLAS $\tau$ program in LHC and future colliders.\\
The computation of the hadronic form factors appearing in $e^+e^-\to2,3$ mesons within this framework can be interesting to include this low-energy description in PHOKHARA and the comparison to the corresponding $CVC$ rotated ones from $\tau$ decays might be helpful in order to understand the size and origin of isospin violations.}

\FullConference{International Workshop on Effective Field Theories: from the pion to the upsilon\\
		 2-6 February 2009\\
		 Valencia, Spain}

\begin{document}

\section{Hadronic decays of the $\tau$ lepton}
Tau Physics has a wealth of motivations \cite{Pich:2008ni}. Here, we take advantage of the fact that, while being the tau a lepton -which within the Standard Model \cite{Glashow:1961tr}, \cite{Pich:2007vu} means that at least half of the process is electroweak, thus clean and under control- its mass ($\sim$ 1.8 GeV) is large enough to connect it to hadrons providing thus a very convenient way to investigate the strong interaction and, particularly how quarks hadronize, a question that remains to be solved from first principles.\\
\indent The decay amplitude for the considered decays may be written as:
\begin{equation} \label{Mgraltau}
\mathcal{M}\,=\,-\frac{G_F}{\sqrt{2}}\,V_{\mathrm{ud/us}}\,\overline{u}_{\nu_\tau}\gamma^\mu(1-\gamma_5)\,u_\tau \mathcal{H}_\mu\,,
\end{equation}
where the strong interacting part is hidden in the hadronic vector, $\mathcal{H}_\mu$:
\begin{equation} \label{Hmugral}
\mathcal{H}_\mu = \bra \left\lbrace  P(p_i)\right\rbrace_{i=1}^n |\left( \mathcal{V}_\mu - \mathcal{A}_\mu\right)  e^{i\mathcal{L}_{QCD}}|0\ket\,.
\end{equation}
\indent Symmetries let us decompose $\mathcal{H}_\mu$ depending on the number of final-state pseudoscalar ($P$) mesons, $n$. For three mesons in the final state, this reads:
\begin{eqnarray} \label{Hmu3m}
\mathcal{H}_\mu = V_{1\mu} F_1^A(Q^2,s_1,s_2) + V_{2\mu} F_2^A(Q^2,s_1,s_2) +
 Q_\mu F_3^A(Q^2,s_1,s_2) + i \,V_{3\mu} F_4^V(Q^2,s_1,s_2)\,,
\end{eqnarray}
and
\begin{eqnarray} \label{VmuQmu}
& & V_{1\mu} \,  = \, \left( g_{\mu\nu} - \frac{Q_{\mu}Q_{\nu}}{Q^2}\right) \,
(p_2 - p_1)^{\nu} ,
\,\,\,\,V_{2\mu} \, = \, \left( g_{\mu\nu} - \frac{Q_{\mu}Q_{\nu}}{Q^2}\right) \,
(p_3 - p_1)^{\nu}\,,\nonumber\\
& & V_{3\mu} \, = \, \varepsilon_{\mu\nu\varrho\sigma}\,p_1^\nu\, \,p_2^\varrho\, \,p_3^{\sigma} ,
 \,\,\,\,Q_\mu \, = \, (p_1\,+\,p_2\,+\,p_3)_\mu \,,\,\,\,\,s_i = (Q-p_i)^2\,.
\end{eqnarray}
\indent $F_i$, $i=1,2,3$, correspond to the axial-vector current ($\mathcal{A}_\m$) while $F_4$ drives the vector current ($\mathcal{V}_\m$). The form factors $F_1$ and $F_2$ have a transverse structure in the total hadron momenta, $Q^\mu$, and drive a $J^P=1^+$ transition. The pseudoscalar form factor, $F_3$, vanishes as $m_P^2/Q^2$ and, accordingly, gives a tiny contribution. This is as far as we can go without model assumptions, that is, it is not yet known how to derive the $F_i$ from $QCD$ \cite{Fritzsch:1973pi}, \cite{Marciano:1977su}.
\section{The K\"uhn-Santamar\'ia Model in TAUOLA}
The most popular parameterization describing these decays is based on Ref.~\cite{Kuhn:1990ad} and it is one of those implemented in the TAUOLA library \cite{Jadach:1993hs}. Despite the good fit it provides for the $2\pi$ and $3 \pi$ modes some of its assumptions may be improved \cite{Roig:2008xt}. The works done in analogy for other decay modes \cite{Finkemeier:1995sr}, however, use some unjustified simplifications that should definitely be corrected \cite{Roig:2008xt}.\\
In order to take full advantage both of the thorough and complete TAUOLA infrastructure and of the very good new experimental data, we believe it is mandatory to use MonteCarlo Generators that incorporate a parameterization for the hadronic matrix elements as close as possible to QCD, in order not to fake the understanding of the measurements. We present in the following a framework capable of accomplishing this task.
\section{Theoretical basis of our approach}
Semileptonic tau decays span the low- and intermediate-energy regions of $QCD$ where it is non-perturbative. As a consequence of Weinberg's Theorem \cite{Weinberg:1978kz}, a valid -and moreover advisable \cite{Georgi:1994qn}- procedure will be a Lagrangian approach including explicitly the active degrees of freedom (the lightest pseudoscalars and resonances) respecting the assumed symmetries. While for the former range of energies there is an unambiguous and successful EFT approach to the problem given by $\chi PT$ \cite{ChPT}, for the latter the situation is much more intricate. In order to build a perturbative approach therein, the inverse of the number of colours was proposed \cite{Nc}. This procedure explained most of light-flavoured meson (and baryon) features \cite{NcToni}.\\
The theory merging the right low-energy behaviour of $\chi PT$ and the 1/$N_C$ expansion of $QCD$ is Resonance Chiral Theory ($R\chi T$) \cite{RChT}. The action determined by symmetries does not share yet $QCD$'s ultraviolet behaviour, so that one needs to demand Brodsky-Lepage conditions \cite{BrodskyLepage} to the appropriate Green functions and associated form factors, which results in restrictions on some combinations of couplings. This has been exploited systematically \cite{RChT}, \cite{GomezDumm:2003ku}, \cite{RuizFemenia:2003hm}, \cite{Cirigliano:2004ue}, \cite{paper}, \cite{RChTHighEnergy} to gain insight in the couplings of the theory and achieve predictivity.\\
Instead of including the infinite number of strictly stable resonances predicted at $LO$ in 1/$N_C$, one would like always to describe Physics in the simplest possible manner, meaning here the Single Resonance Approximation \cite{SRA}: including just the lightest nonet of resonances of given $spin^{parity}$. The historically successful notion of vector meson dominance \cite{VMD} will further reduce the spectrum of exchanged resonances.\\
In any phenomenological application it is mandatory to provide the resonances with a width. Its off-shell behaviour is particularly important for those that are rather wide: $\rho(770)$, $K^*(892)$, $a1(1260)$, etc. This we do following the definition given in Ref. \cite{GomezDumm:2000fz}. Our work \cite{paper} is the first one computing the $a_1$ width in this way.\\
The relevant part of the R$\chi$T Lagrangian is \cite{RChT}, \cite{RuizFemenia:2003hm}, \cite{Cirigliano:2004ue}, \cite{paper}:
\begin{eqnarray} \label{Full_Lagrangian}
 \mathcal{L}_{R\chi T} & = & \frac{F^2}{4}\bra u_\mu u^\mu +\chi_+ \ket\,+\,\frac{F_V}{2\sqrt{2}}\bra V_{\mu\nu} f^{\mu\nu}_+ \ket \,+\, \frac{i \,G_V}{\sqrt{2}} \bra V_{\mu\nu} u^\mu u^\nu\ket \,+\, \frac{F_A}{2\sqrt{2}}\bra A_{\mu\nu} f^{\mu\nu}_- \ket \,+\,\mathcal{L}_{\mathrm{kin}}^V\nonumber\\
& & +\, \mathcal{L}_{\mathrm{kin}}^A \,+\, \sum_{i=1}^{5}\lambda_i\mathcal{O}^i_{VAP} \,+\, \sum_{i=1}^7\frac{c_i}{M_V}\mathcal{O}_{VJP}^i\, +\, \sum_{i=1}^4d_i\mathcal{O}_{VVP}^i \,+\, \sum_{i=1}^5 \frac{g_i}{M_V} {\mathcal O}^i_{VPPP} \, ,
\end{eqnarray}
where all couplings are real, being $F$ the pion decay constant in the chiral limit. The notation is that of Ref.~\cite{RChT}. $P$ stands for the lightest pseudoscalar mesons and $A$ and $V$ for the (axial)-vector mesons. Furthermore, all couplings in the last line are defined to be dimensionless. The subindex of the operators stands for the kind of vertex described, i.e., ${\mathcal O}^i_{VPPP}$ gives a coupling between one Vector and three Pseudoscalars. For the explicit form of the operators in the last line, see \cite{RuizFemenia:2003hm}, \cite{Cirigliano:2004ue}, \cite{paper}.
\section{The axial-form factor and the $a_1$ width: $\tau\to(3\pi,KK\pi)\,\nu_\tau$}
Our framework describes pretty well the two-meson decays of the $\tau$, as shown in the $\pi\pi$ \cite{pipi} and $K\pi$ \cite{Kpi} cases. We focus here on the three meson modes stated above.\\ The proposed expression -that complies \cite{Jorge06} with the $\chi PT$ prediction at $NLO$ \cite{Colangelo:1996hs}- depends on four unrestricted couplings. Short-distance constraints impose relations between them in such a way that only one coupling ($\lambda_0$) remains free. A fit to ALEPH spectral function, branching ratio and integrated structure functions \cite{aleph3pi} yields  $12$ \cite{GomezDumm:2003ku} for it, that is roughly two orders of magnitude larger than the estimate in Ref. \cite{Cirigliano:2004ue} for $\bra VAP\ket$ due to the mild dependence on $\lambda_0$ in this particular mode ($\sim m_\pi^2/Q^2$). Although it is possible to obtain a very good fit of the normalized spectral function with lower values of $\lambda_0$ \cite{Roig:2008xt} satisfying the more general  constraints valid for all three meson processes \cite{Roig:2007yp}, it is not possible to obtain so good branching ratios simultaneously. Upon relaxing Weinberg sum rules letting their predictions vary within some $1/3$ to account for the effect of cutting the infinite tower of resonances present in the $N_C\to \infty$ limit of large-$N_C$ $QCD$ and introducing the effect of the $\rho(1450)$ resonance, we are able to improve the description, as it is illustrated in Fig. 
1. The same setting provides a sensible description of the $\tau\to(KK\pi)\,\nu_\tau$ processes.
\begin{figure}[ht] \label{tau3pi}
  \begin{center}
\centerline{\includegraphics[scale=0.3,angle=-90]{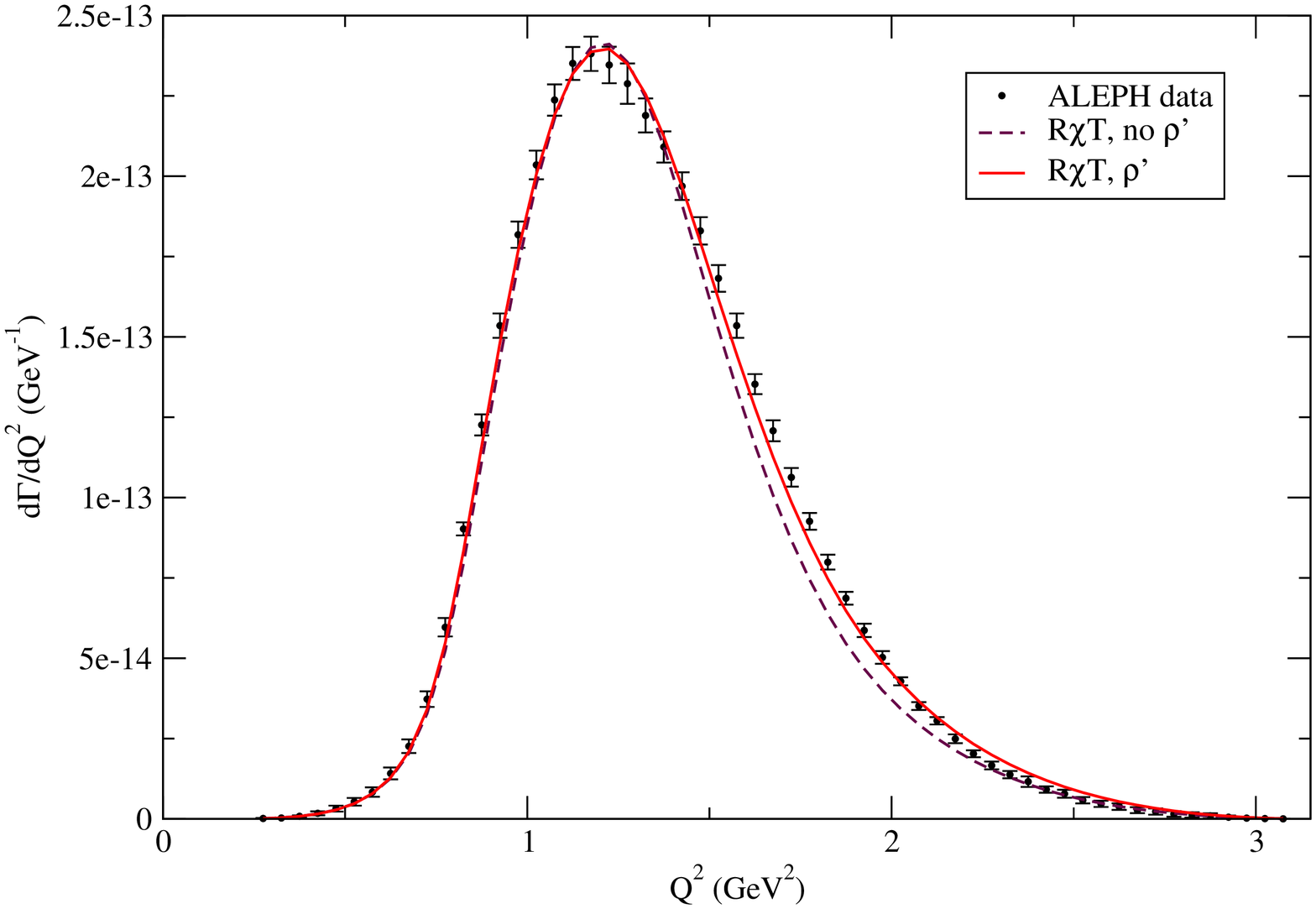} \quad\includegraphics[scale=0.3,angle=-90]{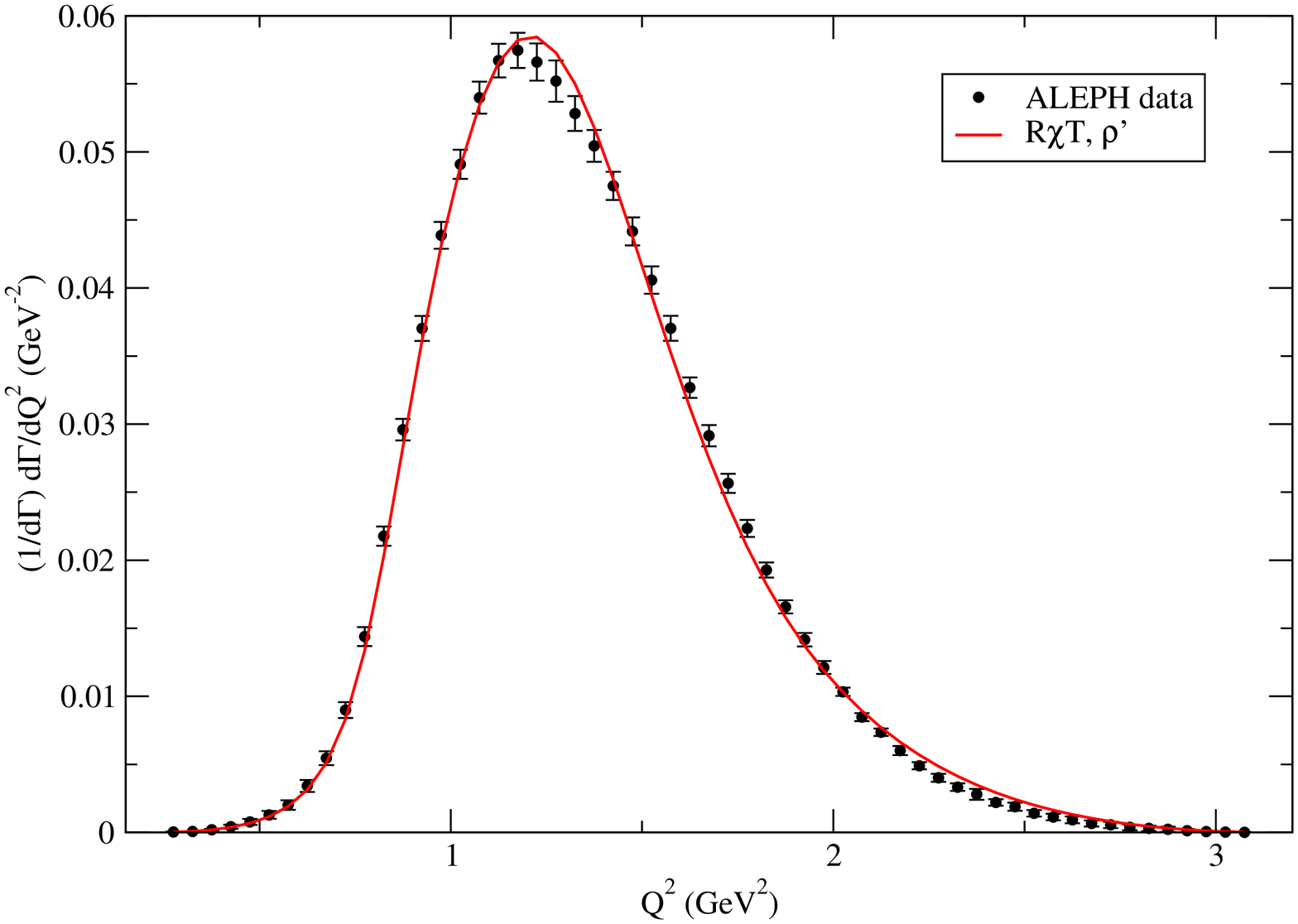}}
\caption{\small{Comparison of our predictions to the ALEPH data \cite{aleph3pi} for the spectral function of $\tau\to (3 \pi)^-\nu_\tau$ (left pane) and its normalized distribution (right pane). The presence of the $\rho' \equiv \rho(1450)$ is necessary to achieve a good description of the data.}}
 \end{center}
\end{figure}
\section{The vector form factor: $\tau\to KK\pi\nu_\tau$ and $\sigma(e^+e^-\to \mathrm{hadrons})$}
After imposing high-energy conditions, we have completed the computation in \cite{RuizFemenia:2003hm} of $\omega \to 3\pi$ including the local piece and have used some of the relations among couplings they obtained. This is not yet enough to bound all couplings entering the vector form factor. However, we could use the branching ratios of the two independent modes to estimate those two remaining. These values have been checked since we could benefit from the nice splitting of the isoscalar and isovector components appearing in $e^+e^-\to KK\pi$ measured by BaBar \cite{babareeKKpi} by radiative return, to use the $SU(2)$-rotated $I=1$ piece to test our outcome. As a result of the procedure, we predict the spectra of all modes and find that vector and axial-vector current are of similar importance in these decay channels both for the decay width and for the shape of the spectrum, as can be seen in Figure 
2.
\begin{figure}[ht] \label{WGmeetinglast}
  \begin{center}
\centerline{\includegraphics[scale=0.3,angle=-90]{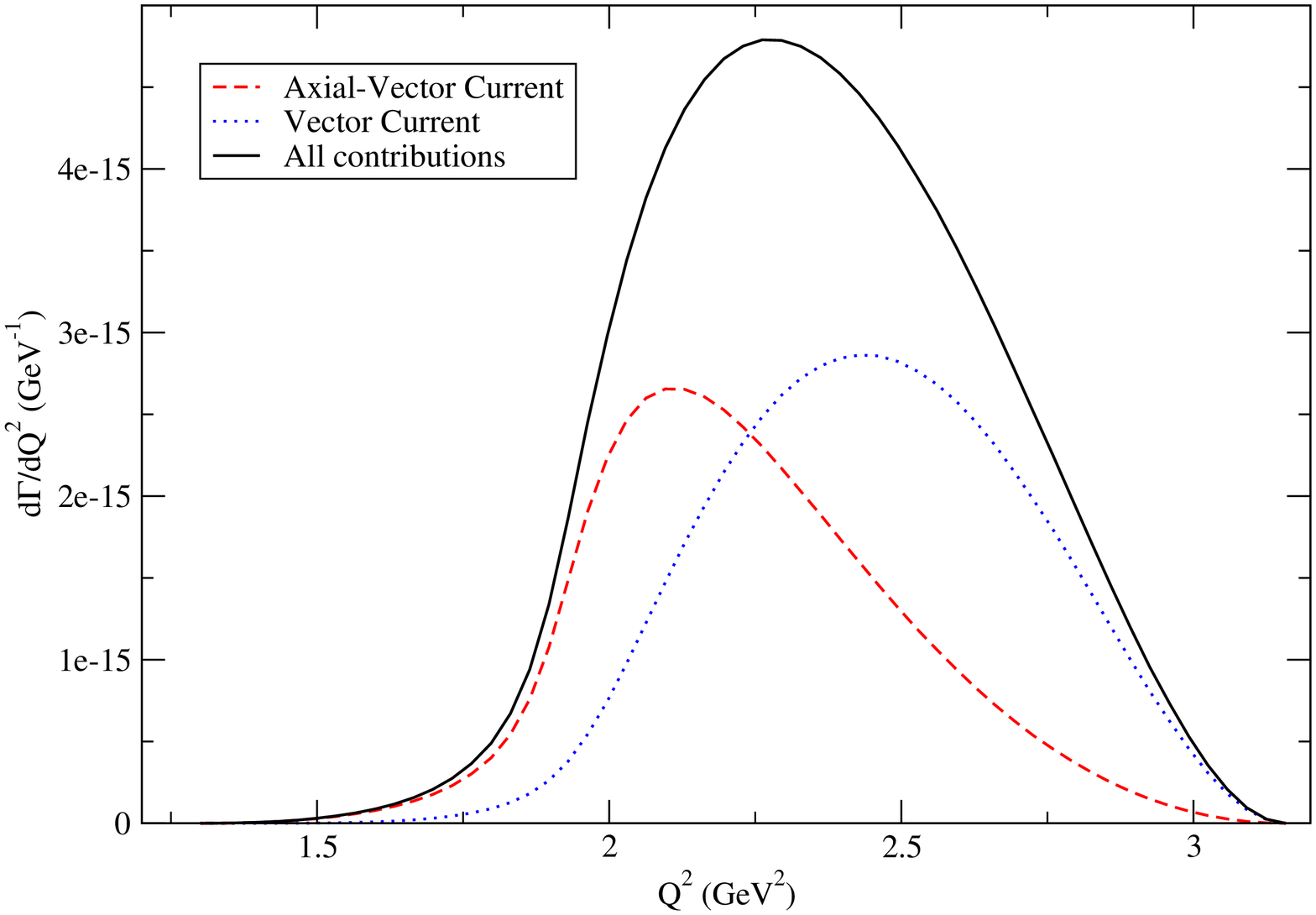} \quad\includegraphics[scale=0.3,angle=-90]{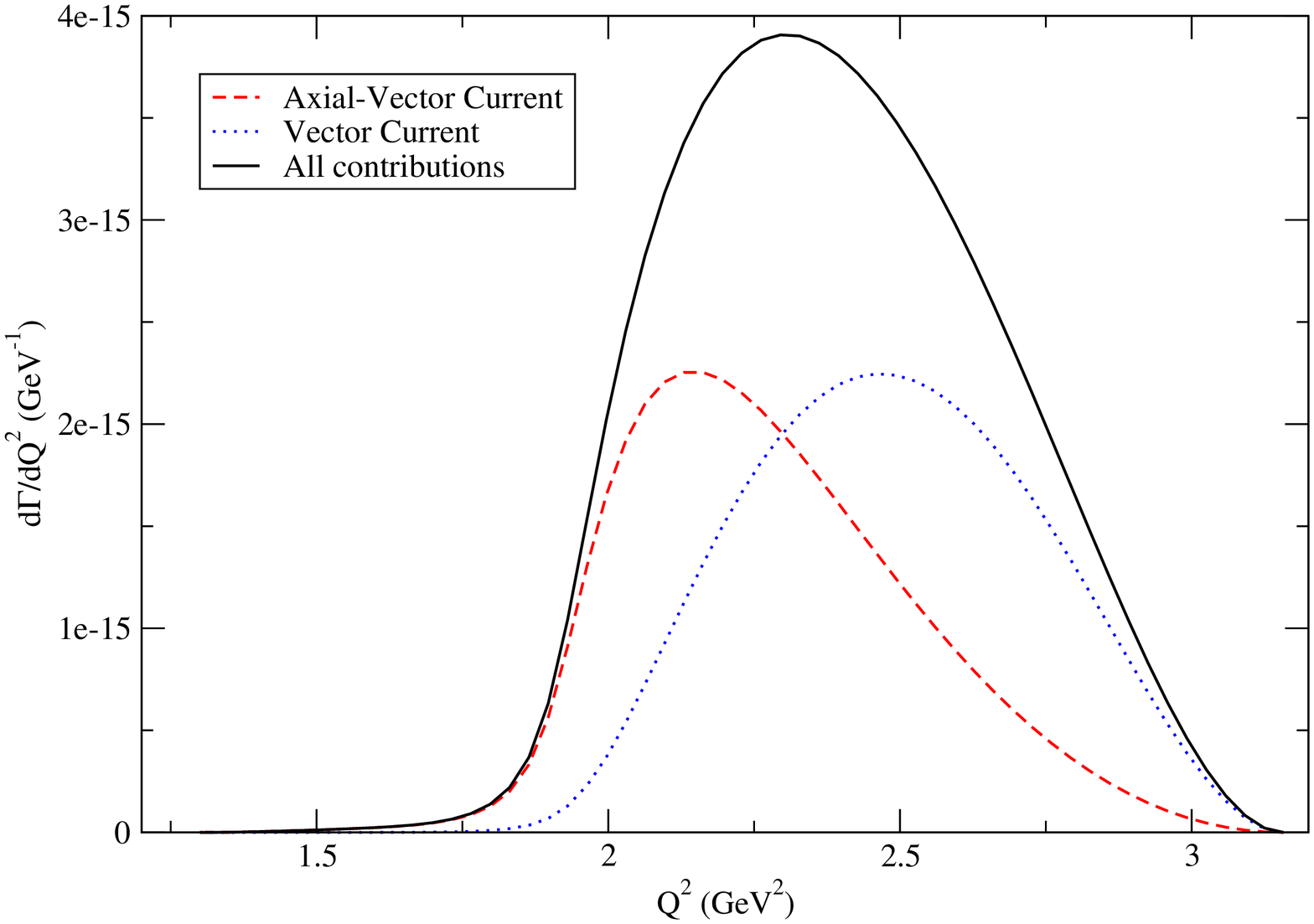}}
\caption{\small{Spectral function of $\tau\to K^+K^-\pi^-\nu_\tau$ (left pane) and $\tau\to K^-K^0\pi^0\nu_\tau$ (right pane). While at lower energies the axial-vector current gives the dominant contribution, at higher energies the vector one gets more importance being responsible of the tail of the curve.}}
 \end{center}
\end{figure}
\section{Conclusions and Outlook}
We have analyzed three meson decays of the $\tau$ within $R\chi T$. We were based on three fundamental pillars of $QCD$: the right low-energy limit, as given by its approximate chiral symmetry, its large-$N_C$ expansion and a Brodsky-Lepage behaviour the form factors must have to ensure their ultraviolet fall-off.\\
We have improved the existing description of the three pion modes by using our computation of  $\Gamma_{a_1}$. We have also worked out the $KK\pi$ channels predicting the spectra and ratio of vector and axial-vector contributions.\\
Our expressions for the vector and axial-vector widths have been implemented successfully in the TAUOLA library and the inclusion of the hadronic matrix elements is in progress \cite{Olga}. This way, the experimental comunity will have as its disposal a way of analysing hadronic decays of the tau that includes as much as possible information from the fundamental theory.\\
We will tackle the other three meson modes in future works. It will be important to achieve a good description of the $K\pi\pi$ channels to help improve the simultaneous extraction of $m_s$ and $V_{us}$ \cite{Gamiz:2004ar}. We also plan to study $e^+e^-\to3$ mesons at low energies what can eventually be used by PHOKHARA \cite{PHOKHARA}.

\end{document}